\documentstyle[prb,aps,multicol]{revtex}
\newcommand{\be}{\begin{eqnarray}}
\newcommand{\ee}{\end{eqnarray}}

\newcommand{\ve}{\varepsilon}
\newcommand{\br}{{\bf r}}
\newcommand{\bq}{{\bf q}}

\begin{document}
\draft
\title{Acoustoelectric  pumping through a ballistic point
contact\\
in the presence of magnetic fields}
\author{O. Entin-Wohlman}
\address{\it School of Physics and Astronomy, Raymond and Beverly Sackler
Faculty of Exact Sciences, \\ Tel Aviv University, Tel Aviv 69978,
Israel\\ }
\author{Y. Levinson }
\address{ Department of Condensed Matter Physics, The Weizmann
Institute of Science, Rehovot 76100, Israel}
\author { P. W\"olfle}
\address{Institut f\"ur Theorie der Kondensierten Materie,
Universit\"at Karlsruhe and\\ Institut {\bf f\"ur} Nanotechnologie,
Forschungszentrum, Karlsruhe 76128 Karlsruhe, Germany}
\date{\today}
\maketitle
\begin {abstract}
The acoustoelectric current, $J$, induced in a ballistic point
contact (PC) by a surface acoustic wave is calculated in the
presence of a perpendicular magnetic field, B. It is found that
the dependence of the current on the Fermi energy in the terminals
is strongly correlated with that of the PC conductance: $J$ is
small at the conductance plateaus, and is large at the steps. Like
the conductance, the acoustoelectric current has the same
functional behavior as in the absence of the field, but with
renormalized energy scales, which depend on the strength of the
magnetic field, $|{\rm B}|$.

\end{abstract}

\begin{multicols}{2}
\section{Introduction}

Acoustoelectric currents induced in a two dimensional electron gas (2DEG)
by a surface acoustic wave (SAW) have been observed in point contacts
(PC's) defined in a GaAs/AlGaAs heterostructure by a split
gate\cite{Sh1,Ta97}. This problem has been analyzed theoretically in
several papers\cite{Sh1,To96,Gu96,Gu98,Ma97,LEW00}. In particular, it has
been realized\cite{LEW00} that the acoustoelectric current in a {\it
ballistic} PC does not simply represent the drag of the electrons by the
SAW, but constitutes an example of a {\it pumping phenomenon}. Pumping in
conducting nanostructures\cite{AG99}, first considered in Ref.
\onlinecite{He91}, is recently a subject of considerable
theoretical\cite{Br98,AA98,SAA00,S00,AAK00}, as well as
experimental\cite{Sw99,Po92,Kou91} interest. The pumping current is
excited not by applying a voltage bias to the nanostructure terminals, but
rather by periodically and adiabatically varying certain properties of the
system, i. e. certain parameters of the system Hamiltonian. In the case of
the  acoustoelectric current,
 the periodic
perturbation is formed by the moving potential profile created by the
acoustic wave, of frequency $\omega_{0}$ and wave vector $q$,
\begin{eqnarray}
\delta{\cal U}({\bf r},t)={\cal A}\cos (\omega_{0} t-{\bf q}{\bf r}).
\end{eqnarray}
As the frequency of the acoustic wave is small, this perturbation can
 be considered
as adiabatic; moreover, the potential $\delta{\cal U}$ can be presented as
$\delta{\cal U}(\br,t)=\lambda_{1}(t)\cos\bq\br+\lambda_{2}(t)\sin\bq\br$
with $\lambda_{1}(t)={\cal A}\cos\omega t$ and $\lambda_{2}(t)={\cal
A}\sin\omega t$ being the `pumping parameters'. This form of the pumping
field is very similar to the one suggested by Thouless\cite{Th83}, in his
pioneering discussion of quantized charge transfer.

This paper is devoted to the calculation of the acoustoelectric
current flowing through a ballistic PC subject to a perpendicular
magnetic field due to a SAW propagating along the PC. We employ a
model of non-interacting electrons. Our main result is that the
acoustoelectric current dependence on the Fermi energy in the
leads in the presence of the magnetic field ${\rm B}$ has the same
functional form as for ${\rm B}=0$, with the energy scales (or
alternatively, the length scales) which characterize the PC, being
renormalized and replaced by field dependent quantities. (This
feature is known for the quantum conductance of the PC
\cite{F87,Bu90}.)

The derivation of the acoustoelectric current, in the presence of a
magnetic field, requires quite a lengthy calculation. We therefore have
organized the paper as follows. The next section describes the formulation
of the problem, and gives the final result for the field-dependent
acoustoelectric
current. We discuss there its dependence on the Fermi energy of the
terminals, on the magnetic field and on the wave vector of the SAW.
 Section III includes the formal derivation of a
time-dependent current, induced by a weak time-dependent perturbation, in
a nanostructure of a general geometry, subject to a magnetic field, and (a
possible) dc bias. The formalism is based on time-dependent scattering
states\cite{wolfle}. At the end of that section we obtain the {\it pumped}
acoustoelectric
current, and relate its form to the  formula for the pumping current
 derived by Brouwer\cite{Br98}. The expression
for the acoustoelectric current necessitates the knowledge of the full
scattering matrix of the PC, in the presence of the magnetic field.  We
carry out this calculation for a saddle-point potential
in Sec. IV, and derive there the result reported in Sec. II.

\section{Formulation and results}
\label{results}

A quite realistic model for the confining potential of a PC in the plane
of a 2DEG is the saddle-point potential\cite{LEW00,Bu90,LLS}
\begin{eqnarray}
 U(x,y)&=&\frac{1}{2md^{2}}\Bigl
(-\frac{x^{2}}{L^{2}}+\frac{y^{2}}{d^{2}}\Bigr ).
 \label{spp}
\end{eqnarray}
Here $m$ is the electron mass, $L$ is the length of the PC and $d$ is its
width ($\hbar=1$). For $L\gg d$ this potential corresponds to a waveguide
in the $x$-direction with parabolic walls (at $|x|\lesssim L$) adjusted to
horns (at $|x|\gtrsim L$) with opening angle $d/L$. These horns represent
the left and right terminals at $x=\mp \infty$. The electronic states in
this confining potential are labeled by the energy $E$ and the transverse
channel number $n$. The transmission amplitude of an electron with energy $E$
propagating in the $n$th channel is
\begin{eqnarray}
t(\varepsilon_{n})=\Bigl (1+e^{-2\pi\varepsilon_{n}}\Bigr
)^{-1/2},\label{transco}
\end{eqnarray}
where $\varepsilon_{n}$ is the (dimensionless) energy for the longitudinal
propagation,
\begin{eqnarray}
\label{energ}
\varepsilon_{n}=(E-E_{n})/\delta ,\ \ E_{n}=\Delta (n+1/2),\label{ve}
\end{eqnarray}
with
\begin{eqnarray}
\delta =1/mLd ,\ \ \Delta =1/md^{2},\label{escales}
\end{eqnarray}
being the two energy scales of the saddle point potential.
The energy $E_{n}$ is the
threshold for the propagation in the $n$th channel: modes having a large
positive $\varepsilon_{n}$ are propagating ($t$ is close to 1), while
modes with large negative $\varepsilon_{n}$ are evanescent ($t$ is
exponentially small).

The motion of an electron through a saddle-point potential, when a
constant magnetic field ${\rm B}$
is applied perpendicularly to the $\{x,y\}$-plane,
has been considered by Fertig and Halperin \cite{F87}. Remarkably enough,
they find the same form, Eqs.  (\ref{transco}) and (\ref{ve}), for the
transmission coefficient, where now the channel index $n$ refers to the
Landau level to which the edge state belongs, and the two energy scales
(\ref{escales}) are modified, to be $\delta_{\rm B}$ and $\Delta_{\rm B}$
[see Eq. (\ref{Ham}) below]. For example, at strong magnetic fields such
that $\omega_{\rm B}\equiv |e|{\rm B}/mc\gg\Delta$, then $\Delta_{\rm
B}\simeq \omega_{\rm B}$ while $\delta_{\rm B}\simeq
\Delta\delta/\omega_{\rm B}$. It therefore follows that, (as in the
absence of the field), the conductance (at zero temperature, in units of
$e^{2}/h$) is\cite{Bu90} ${\cal G}=\sum_{n}t^{2}(\ve_{n})$, with $E$ [see
Eq. (\ref{ve})] being the the Fermi energy in the terminals, $E_{F}$. When
$\Delta_{\rm B}\gg \delta_{\rm B}$ the conductance as function of $E$ is a
step-like function, with plateaus of width $\Delta_{\rm B} $ and steps of
width $\delta_{\rm B} $. Increasing the magnetic field makes the intervals
between the steps larger and the width of the steps smaller.

In the following sections we calculate the acoustoelectric
current, in the presence of a magnetic field normal to the 2DEG
plane, and find that it consists of the sum over the contributions
from the various  channels $n$,
\begin{eqnarray}
J=J_{0}\sum_{n}F(\varepsilon_{n},p).\label{Jfinal}
\end{eqnarray}
Here, $J_{0}$ is the nominal current,
\begin{eqnarray}
J_{0}=2e\omega_{0}\frac{|{\cal A}|^{2}}{\delta_{\rm
B}^{2}},\label{nominal}
\end{eqnarray}
 $\varepsilon_{n}$ is defined as in Eq. (\ref{energ})
with $E=E_{F}$, and $p\ $  gives
the SAW wave vector in dimensionless units,
\begin{eqnarray}
p=aq, \ \ a=\Bigl (\frac{1}{2m\delta_{\rm
B}}\frac{\Delta^{2}+\delta^{2}_{\rm B}}{\Delta_{\rm B}^{2}+\delta_{\rm
B}^{2}}\Bigr )^{1/2}.\label{p}
\end{eqnarray}
At zero magnetic field, $a=(Ld/2)^{1/2}$, while at strong fields it
tends to $\ell_{\rm B}(L/2d)^{1/2}$, where $\ell_{\rm B}=(c/|e|{\rm B})^{1/2}$
is the magnetic length.
The function $F(\varepsilon,p)$, given in Eq. (\ref{Ffinal}) below,
is the same one which appears in the calculations of the zero-field
acoustoelectric current \cite{LEW00}. This function
is exponentially small well above and below the threshold, where
$|\varepsilon|\gg 1$. Near the threshold, where $|\varepsilon|\lesssim 1$,
this function  for $p\gg 1$ displays damped oscillations as function of
$p$ with a period $\Delta p\simeq 1$, and tends to zero at $p\to 0$,

\begin{eqnarray}
\label{F}
 F(\ve,p)&=&2\pi e^{-\pi\ve}t^{3}(\ve)c(\ve)\;{\rm erf}
 \left(\frac{p}{\sqrt{2\sigma}}\right) ,\  p\ll
1,\nonumber\\
F(\ve,p)&=&4\pi t^{2}(\ve)\frac{e^{-\sigma p^{2}}}{p^2}
\cos^2\left({p^2\over
2}-{\pi\over 4}-\gamma_{\ve}\right),\; p\gg 1.
\end{eqnarray}
Here  $\gamma_{\ve}=2\ve\ln p+\arg\Gamma(1/2-i\ve)$, $c(\ve)\simeq
1$ and $\sigma\simeq d/L$ is a screening parameter. It is
introduced phenomenologically in order to  model the screening of
the SAW potential in the wide banks of the PC \cite{LEW00}. Since
the expression (\ref{Jfinal}) for the acoustoelectric current is
the same function  as in the absence of the magnetic
field\cite{LEW00},  but with modified, field-dependent, arguments,
we  may discuss its properties borrowing from our\cite{LEW00}
previous results.

As in the ${\rm B}=0$ case, the modes whose  thresholds $E_{n}$
are far from the Fermi energy $E_{F}$ do not contribute to the
current. This is expected for the evanescent modes; for the
propagating ones this seems to be ``counter-intuitive". It means
that in a free channel with almost no reflection, the
acoustoelectric current is zero. As a result, the current is large
only when the Fermi energy is close to one of the thresholds,
$|E_F-E_{n}|\lesssim \delta_{\rm B}$. Hence, as in the ${\rm B}=0$
case, the dependence of the acoustoelectric current on the Fermi
energy closely follows that of the conductance. Since $\delta_{\rm
B}<\delta$, increasing the magnetic field reduces the domains
where the current is strong, the squeezing factor at strong fields
being  $\omega_{{\rm B}}/\Delta$.
 This is in agreement with the expectation that pumping is quenched
by a magnetic field due to the chirality of the electron states.
Less expected is that, on
 the other hand, the magnetic field enhances the magnitude of
the nominal current $J_{0}$, Eq. (\ref{nominal}), since
$\delta_{\rm B}<\delta$. This means that while away from the
threshold the current is reduced,  the field enhances the current
at the threshold, the enhancement factor at strong fields being
$(\omega_{{\rm B}}/\Delta)^2$. The field changes also the
oscillation period in $q$, which is of the order $a^{-1}$ and
hence increases with the field. At strong fields $a/a_{\rm
B=0}=d/\ell_{\rm B}$. We also note that the acoustoelectric
current in a saddle-point potential PC is invariant with respect
to ${\rm B}\rightarrow -{\rm B}$, in agreement with the up-down
symmetry discussed in Ref.\onlinecite{AAK00}.

The scaling of the characteristic energies with magnetic field is
unique to the saddle-point potential and is not a generic feature.
However, since the saddle-point potential is a good approximation
to the actual confining potential of a PC in a 2DEG, one  expects
that at least the trends predicted by the scaling will be valid
for a realistic confining potential.

\section{Time-dependent response}

The acoustoelectric current, driven by a SAW, is an example of a
dc current induced by an ac perturbation. Quite generally, such a
current can be obtained using the concept of {\it time-dependent
scattering states} \cite{wolfle}. In this section we use this
formulation to derive the expression for the current flowing in a
multi-terminal
 ballistic nanostructure of a
complicated geometry, which may be subject to a dc bias, in the
presence of a magnetic field. The result is given in Eq.
(\ref{Jbeta}) below, and is valid up to second order in the
strength of the ac perturbation. We then specify to the ``pumping"
conditions, namely, a dc current induced in an unbiased
nanostructure, and show that the result is a generalization of
Brouwer's formula \cite{Br98} for the pumped charge, for the case
of spatially distributed pumping parameters.

In order to use the time-dependent scattering state formalism, one writes
the current density operator, ${\bf j}({\bf
r},t)=-(ie/2m)\Psi^{\dagger}({\bf r},t){\tilde\nabla}\Psi ({\bf
r},t)+h.c.$, with ${\tilde\nabla}\equiv \nabla -(ie/c){\bf A}({\bf r},t)$
(where ${\bf A}$ denotes the vector-potential) representing the electron field
operator $\Psi$ with the time-dependent scattering states \cite{wolfle},
\begin{eqnarray}
\Psi ({\bf r},t)=\int\frac{dE}{2\pi}\sum_{\alpha n}a_{\alpha
n}(E)\chi_{\alpha n}(E|{\bf r},t).\label{psi}
\end{eqnarray}
The state $\chi_{\alpha n}(E|{\bf r},t)$ is a solution of the
Schr\"{o}dinger equation with a time-dependent Hamiltonian
\begin{eqnarray}
{\cal H}({\bf r},t)={\cal H}_{0}({\bf r})+\delta{\cal U}({\bf r},t),
\end{eqnarray}
excited by an electron coming with energy $E$ from channel $n$ of terminal
$\alpha$. The solutions are normalized to unit incoming flux.
Although the time-dependent scattering state is labeled with the energy $E$ of
the incoming wave, it contains components with energies $E'\neq E$ since
the Hamiltonian is time-dependent. In (\ref{psi}), the operator $a_{\alpha
n}(E)$ annihilates an electron of energy $E$ in the $n$th channel of lead
$\alpha$. The thermal average of these operators is defined, in general,
by the temperature and the chemical potential of the terminal connected to
the lead, i. e.,
\begin{eqnarray}
\langle a^{\dagger}_{\alpha n}(E)a_{\alpha 'n'}(E')\rangle =2\pi\delta
(E-E')\delta_{\alpha n,\alpha 'n'}f_{\alpha}(E),\label{thermal}
\end{eqnarray}
where $f_{\alpha}(E)=(e^{(E-\mu_{\alpha})/k_{\rm B}T}+1)^{-1}$ is
the Fermi distribution with the chemical potential $\mu_\alpha$ in
terminal $\alpha$. The next step is to find the
average of the current density operator. Then, the total current entering
terminal $\beta $ is given by
\begin{eqnarray}
J_{\beta }(t)=\int_{C_{\beta}}d{\bf r}\langle {\bf j}({\bf r},t)\rangle
|_{{\bf r}\rightarrow\infty\beta},\label{thermal1}
\end{eqnarray}
where the notation ${\bf r}\rightarrow\infty\beta$ means that ${\bf r}$
approaches infinity in lead $\beta $, whose cross section is denoted
$C_{\beta}$. Therefore, it is sufficient to obtain the states
$\chi_{\alpha n}$ for ${\bf r}\rightarrow\infty\beta$.

In the case at hand,  ${\cal H}_{0}$ is the Hamiltonian of the PC
and $\delta{\cal U}({\bf r},t)=\int d\omega \exp(-i\omega t)
\delta{\cal U}({\bf r},\omega )$ is the potential created by the
SAW. A straightforward calculation yields that, to second order in
the perturbation $\delta{\cal U}$, the time-dependent scattering
state is
\begin{eqnarray}
&&\chi_{\alpha n}(E|{\bf r},t)=e^{-iEt}\Bigl [\chi_{\alpha n}(E|{\bf r})
\nonumber\\ &+&\int d{\bf r}'d\omega 'e^{-i\omega 't}\delta{\cal U}({\bf
r}',\omega ') \Bigl ( G(E' |{\bf r},{\bf r}')\nonumber\\ &+&\int d{\bf
r}''d\omega ''G(E''|{\bf r}, {\bf r}'')\delta{\cal U}({\bf r}'',\omega '')
e^{-i\omega ''t}\nonumber\\ &&\times G(E'|{\bf r}'',{\bf r}')\Bigr )
\chi_{\alpha n}(E|{\bf r}')\Bigr ],\label{scs}
\end{eqnarray}
where we have denoted $E'\equiv E+\omega '$ and $E''\equiv E+\omega
'+\omega ''$. Here $G$ is the Green function of the Hamiltonian ${\cal
H}_{0}$, and $\chi_{\alpha n}(E|{\bf r})$ is the scattering state of
${\cal H}_{0}$, i. e., in the absence of the time-dependent perturbation.
The asymptotic behavior of the latter at
${\bf r}\rightarrow\infty\beta$ was found in \cite{L00},
\begin{eqnarray}
\chi_{\alpha n}(E|{\bf r})|_{{\bf
r}\rightarrow\infty\beta}&=&\delta_{\alpha\beta}w^{-}_{\alpha n}(E|{\bf
r})\nonumber\\ &+&\sum_{m}w^{+}_{\beta m}(E|{\bf r})S_{\beta m,\alpha
n}(E),
\end{eqnarray}
where $S_{\beta m,\alpha n}(E)$ is the scattering matrix element from
$\alpha n$ to $\beta m$ for the Hamiltonian ${\cal H}_{0}$, and $w^{-}$,
$w^{+}$, are the incoming and the outgoing waves, respectively. The
asymptotic behavior of the time-dependent scattering state (\ref{scs}) is
derived from this relation and the asymptotic behavior of the Green
function  \cite{L00} ,
\begin{eqnarray}
G(E|{\bf r},{\bf r}')|_{{\bf
r}\rightarrow\infty\beta}=-i\sum_{m}w^{+}_{\beta m}(E|{\bf
r})\hat{\chi}_{\beta m}(E|{\bf r}'),
\end{eqnarray}
where the ``hat" indicates that the magnetic field is inverted.
One finds
\begin{eqnarray}
&&\chi_{\alpha n}(E|{\bf r},t)|_{{\bf
r}\rightarrow\infty\beta}=e^{-iEt}\Bigl [\delta_{\alpha\beta}w^{-}_{\alpha
n}(E|{\bf r})\nonumber\\ &+&\sum_{m}\Bigl (S_{\beta m,\alpha
n}(E)w^{+}_{\beta m}(E|{\bf r})\nonumber\\ &+&\int d\omega 'e^{-i\omega
't}S^{(1)}_{\beta m,\alpha n}(E',E)w^{+}_{\beta m}(E'|{\bf r})\nonumber\\
&+&\int d\omega' d\omega ''e^{-i(\omega '+\omega '')t}\nonumber\\ &&\times
S^{(2)}_{\beta m,\alpha n}(E'',E',E)w^{+}_{\beta m}(E''|{\bf r})\Bigr
)\Bigr ],
\end{eqnarray}
with
\begin{eqnarray}
&&S^{(1)}_{\beta m,\alpha n}(E' ,E)=\nonumber\\ &-&i\int d{\bf r}
\hat{\chi}_{\beta m}(E'|{\bf r})\delta{\cal U}({\bf r},\omega
')\chi_{\alpha n}(E|{\bf r}),\nonumber\\ &&S^{(2)}_{\beta m,\alpha
n}(E'',E' ,E)=\nonumber\\ &-&i\int d{\bf r}d{\bf r}'\hat{\chi}_{\beta
m}(E''|{\bf r}')\delta{\cal U}({\bf r}',\omega '')\nonumber\\ &&\times
G(E'|{\bf r}',{\bf r})\delta{\cal U} ({\bf r},\omega ')\chi_{\alpha
n}(E|{\bf r}).\label{scat}
\end{eqnarray}

Upon inserting this expression into Eq. (\ref{thermal1}) for the current,
using Eq. (\ref{thermal}), one finds that the current is given in terms of
current density matrix elements for waves $w^{\pm}_{\beta m}$.
Only diagonal matrix elements contribute, and using the normalization
of $w^{\pm}_{\beta m}$  to a unit flux, one obtains
\begin{eqnarray}
J_{\beta}(t)&=&-e\int\frac{dE}{2\pi}\Bigl
\{\sum_{n}f_{\beta}(E)\nonumber\\ & -&\sum_{\alpha
mn}f_{\alpha}(E)|\tilde{S}_{\beta m,\alpha n}(E,t)|^{2}\Bigr
\},\label{Jbeta}
\end{eqnarray}
where a time-dependent scattering matrix, $\tilde{S}$, has been
introduced,
\begin{eqnarray}
&&\tilde{S}_{\beta m,\alpha n}(E,t)=S_{\beta m,\alpha n}(E)\nonumber\\
&+&\int d\omega 'e^{-i\omega 't}S_{\beta m,\alpha n}^{(1)}(E'
,E)\nonumber\\ &+&\int d\omega 'd\omega ''e^{-i(\omega '+\omega
'')t}S^{(2)}_{\beta m,\alpha n}(E'',E',E).
\end{eqnarray}

The result (\ref{Jbeta}) gives the current induced in the system
by a weak time-dependent perturbation to second-order in the
perturbation. (The first order contribution to ``internal
potential" in terms of the scattering matrix was calculated in
Ref. \onlinecite{pretre}). It hence may serve to obtain a general
ac response in the presence of a dc bias. In the following
discussion, however, we confine ourselves to the case of a dc
response, in the absence of a dc bias.

To obtain the {\it dc, pumped} current we now  assume that the
nanostructure is unbiased, i. e. all terminals have the same chemical
potential, and select from the general expression only the
time-independent terms. The resulting expression is
\begin{eqnarray}
J^{dc}_{\beta}&=&e\int\frac{dE}{2\pi}\int d\omega \Bigl (f(E)-f(E+\omega
)\Bigr ) \nonumber\\ &&\times\sum_{\alpha mn}| S^{(1)}_{\beta m,\alpha
n}(E' ,E)|^{2}.\label{form}
\end{eqnarray}
Note that the second order contribution to the time dependent
scattering matrix $S^{(2)}$ does not enter  this expression. To
derive it, we have used the following properties  of the
scattering states and the Green function \cite{L00},
\begin{eqnarray}
\sum_{\alpha n}S^{\ast}_{\beta m,\alpha n}\chi_{\alpha
n}\equiv\sum_{\alpha n}\hat{S}^{\ast}_{\alpha n,\beta m}\chi_{\alpha
n}=\hat{\chi}^{\ast}_{\beta m},\label{property}
\end{eqnarray}
and
\begin{eqnarray}
&&G(E|{\bf r}_{2},{\bf r}_{1})-G^{\ast}(E|{\bf r}_{1},{\bf
r}_{2})\nonumber\\
 &=&-i\sum_{\alpha n}\chi_{\alpha n}(E|{\bf
r}_{2})\chi^{\ast}_{\alpha n}(E|{\bf r}_{1}).
\end{eqnarray}

 The form (\ref{form}) of the
pumping current can be presented as a generalization of the
adiabatic pumping formula given in Ref. \onlinecite{Br98}. One
notes that $S^{(1)}$, Eq. (\ref{scat}), in the adiabatic
approximation is just the correction $\delta S$ to the scattering
matrix due to the small perturbation $\delta{\cal U}$ (see also
Ref. \onlinecite{pretre}). In terms of scattering states this
correction is
\begin{eqnarray}
\frac{\delta S_{\beta m,\alpha n}(E)}{\delta{\cal U}({\bf r},\omega
)}=-i\hat{\chi}_{\beta m}(E|{\bf r})\chi_{\alpha n}(E|{\bf r}),
\end{eqnarray}
and consequently,
\begin{eqnarray}
&&J^{dc}_{\beta}=e\int \frac{dE}{2\pi}\int d\omega \omega
\left(-\frac{\partial f(E)}{\partial E}\right)
\nonumber\\ &&\times\int d{\bf r}_{1}d{\bf
r}_{2}\delta{\cal U}({\bf r}_{1},-\omega )\delta{\cal U}({\bf
r}_{2},\omega )\nonumber\\ &&\times\sum_{\alpha mn}\frac{\delta S_{\beta
m,\alpha n} (E)}{\delta{\cal U}({\bf r}_{2},\omega )} \frac{\delta
S^{\ast}_{\beta m,\alpha n}(E)}{\delta{\cal U}({\bf r}_{1},-\omega )}.
\end{eqnarray}
It follows that the above derivation gives the pumping current for a
multi-terminal device, in the presence of a magnetic field, which is
pumped by a weak modulation of the ``distributed parameters" $\delta{\cal
U}({\bf r})$.

\section{The acoustoelectric current}

When the time-dependent potential is created by a SAW moving along the
$x$-direction, one has
\begin{eqnarray}
\delta{\cal U}({\bf r},\omega )={\cal A}e^{iqx}\delta (\omega
-\omega_{0})+{\cal A}^{\ast}e^{-iqx}\delta (\omega +\omega_{0}).
\end{eqnarray}
Then introducing this into  $S^{(1)}$ given by Eq. (\ref{scat})
one finds that at zero temperature, the current pumped by the SAW
is given by
\begin{eqnarray}
J^{dc}_{\beta}=e\omega_{0}|{\cal A}|^{2}\sum_{\alpha mn}\Bigl
(|\tilde{M}^{q}_{\beta m ,\alpha n}|^{2}-|\tilde{M}^{-q}_{\beta m,\alpha
n}|^{2}\Bigr ),     \label{JSAW}
\end{eqnarray}
with
\begin{eqnarray}
\tilde{M}^{q}_{\beta m,\alpha n}(E)=\int d{\bf r}\hat{\chi}_{\beta
m}(E|{\bf r})e^{iqx}\chi_{\alpha n}(E|{\bf r}),\label{integrals}
\end{eqnarray}
where now $E=E_{F}$.

The evaluation of the integrals (\ref{integrals}) requires the solution of
the scattering problem of a saddle-point potential in the presence of a
magnetic field. This may be accomplished by choosing the symmetric gauge
for the vector potential, ${\bf A}=({\rm B}/2)\{-y,x,0\}$, and changing
the variables $x,p_{x}$ into $\xi ,s$, as has been done in Ref.
\onlinecite{F87}.  As discussed there, $\xi $ corresponds to the
$x$-coordinate of the guiding center, while $s$ describes the cyclotron
motion about that center. The advantage of the $\{\xi ,s\}$-representation
is that the Hamiltonian, which includes the magnetic field, takes again
the same form of a saddle-point potential,
\begin{eqnarray}
{\cal H}_{0}&=&\delta_{B}\Bigl
(-\frac{\partial^{2}}{\partial\xi^{2}}-\frac{1}{4}\xi^{2}\Bigr )
+\frac{1}{2}\Delta_{B}\Biggl (-\frac{\partial^{2}}{\partial
s^{2}}+s^{2}\Biggr )\label{ham0}
\end{eqnarray}
 but with magnetic field dependent energy scales:
\begin{eqnarray}
\delta_{B}&=&2\sqrt{\gamma^{2}-\lambda_{-}^{2}}, \ \ \Delta_{B}
=2\sqrt{\lambda_{+}^{2}-\gamma^{2}}.\label{Ham}
\end{eqnarray}
That is, in the $\{\xi ,s\}$-representation, the variables are separated.
The parameters of this transformation are given by
\begin{eqnarray}
 \left[\begin{array}{c}x\\p_{y}\end{array}\right]=
 \left[\begin{array}{cc}\ell e^{-\theta_{1}}\cos\phi&\ell\sqrt{2}
e^{\theta_{2}}\sin\phi\\
\displaystyle\frac{1}{2\ell}e^{-\theta_{1}}\sin\phi&
-\displaystyle\frac{1}{\sqrt{2}\ell}e^{\theta_{2}}\cos\phi
\end{array}\right]
\left[\begin{array}{c}\xi\\s\end{array}\right].\label{aij}
\end{eqnarray}
Here,
\begin{eqnarray}
\ell =(2m\lambda)^{-1/2},\ &&\
 \tan 2\phi=-\omega_{B}/4\gamma,\nonumber\\
 {\rm tanh} 2\theta_{1}=
-\lambda_{-}/\gamma,\ &&\ {\rm tanh} 2\theta_{2}=\gamma/\lambda_{+},
\end{eqnarray}
where
\begin{eqnarray}
\lambda^{2}&=&\frac{1}{4}\omega_{B}^{2}+\frac{1}{2}(\Delta^{2}-\delta^{2}),
\ \ \lambda\gamma =\frac{1}{8}(\Delta^{2}+\delta^{2}),\nonumber\\
\lambda_{\pm}&=&\frac{1}{2}\lambda\pm \sqrt{\gamma^{2}+\Bigl
(\frac{1}{4}\omega_{B}\Bigr )^{2}}.
\end{eqnarray}
 The scattering solutions of Hamiltonian
(\ref{ham0}), of energy $E$, are labeled by $\alpha n$, where
$\alpha =1,2$ correspond to the left and right terminals,
 and and $n$ is the channel index.
$\chi_{1n}(E|\xi ,s)$ describes an electron entering the PC from
$x=-\infty$, and $\chi_{2n}(E|\xi ,s)$ corresponds to an electron
coming from $x=+\infty$.   In the presence of the magnetic field
\cite{F87} $\chi_{1n}$ describes an electron entering from the
upper left corner (assuming ${\rm B} >0$) of the $\{x,y\}$-plane,
while $\chi_{2n}$ belongs to an electron coming from the right
lower corner. We write the scattering states in the form
\begin{eqnarray}
\chi_{\alpha n}(E|\xi ,s)=C\Phi_{n}(s)\chi_{\alpha n}(\varepsilon_{n}|\xi
),\ \ \alpha =1,2.\label{fullss}
\end{eqnarray}
Here $C$ is the normalization constant, chosen to be real (to be
determined below), and $\Phi_{n}$ are the
eigenfunctions of the $s$-part of the Hamiltonian (\ref{Ham}),
i.e. harmonic oscillator wave functions, corresponding to
eigenenergies $E_{n}=\Delta_{\rm B}(n+1/2)$ and
 normalized to unity,
$\int_{-\infty}^{\infty}ds\Phi^{2}_{n}(s)=1$. Then $\chi_{\alpha
n}(\varepsilon_{n}|\xi )$ are the scattering states of the
one-dimensional problem\cite{LLS}, defined by the $\xi$-part of
the Hamiltonian (\ref{Ham}), with (dimensionless) energies
$\varepsilon_{n}$,
\begin{eqnarray}
\chi_{1,2}(\varepsilon_{n} |\xi )= -it(\varepsilon_{n} ){\rm{\bf
E}}(-\varepsilon_{n} ,\pm\xi ),   \label{weber}
\end{eqnarray}
and ${\rm {\bf E}}$ is the complex Weber (parabolic cylinder)
function, as defined in Ref. \onlinecite{Ab64}.

The integrals (\ref{integrals}) are not matrix elements because
the functions $\chi$ and $\hat{\chi}$ belong to different
Hamiltonians, one with ${\rm B}$ and the other with $-{\rm B}$. As
can be appreciated from the above analysis of the saddle-point
potential in the presence of a magnetic field, the transformation
(\ref{aij}) turns the scattering problem formally into the one in
the absence of the field, but in the mixed representation
$\{$momentum, coordinate$\}$. It follows that in order to perform
the integrals (\ref{integrals}), it is convenient to convert them
first into matrix elements, which are then straightforwardly
transformed in accordance with (\ref{aij}). This can be done by
using the relations (\ref{property}) to obtain
\begin{eqnarray}
\tilde{M}^{q}_{\beta m,\alpha n}(E)=\sum_{\beta 'm '}S_{\beta m,\beta
'm'}(E)M^{q}_{\beta 'm',\alpha n}(E),
\end{eqnarray}
with
\begin{eqnarray}
M^{q}_{\beta m,\alpha n}(E)&=&\int d{\bf r}\chi^{\ast}_{\beta m}(E|{\bf
r})e^{iqx}\chi_{\alpha n}(E|{\bf r})\nonumber\\ &&\equiv\langle \beta
m|e^{iqx}|\alpha n\rangle .\label{matrixele}
\end{eqnarray}
Hence, the calculation of the acoustoelectric current is reduced to (i)
obtaining the scattering matrix elements (including their phases, which
are not derived in Ref. \onlinecite{F87}); (ii) normalizing the scattering
states (\ref{fullss}) to a unit flux; and (iii) evaluating the matrix
elements (\ref{matrixele}).

(i) Because of the variable separation in the Hamiltonian (\ref{Ham}), the
scattering matrix does not mix the modes $n$,
\begin{eqnarray}
S_{\alpha n,\beta m}(E)=\delta_{nm}S_{\alpha\beta}(\ve_{n}),  \label{1dme}
\end{eqnarray}
where $S_{\alpha\beta}(\ve_{n})$ is the 2$\times$2 scattering matrix for the
one-dimensional problem defined by the $\xi$-part.
The absolute values of these matrix elements have been
derived in Ref. \onlinecite{F87}: $
|S_{12}(\ve_{n})|=|S_{21}(\ve_{n})|=t(\ve_{n})$, and
$|S_{11}(\ve_{n})|=|S_{22}(\ve_{n})|=r(\ve_{n})\equiv
e^{-\pi\ve_{n}}t(\ve_{n}).$ To find the {\it phases} of the matrix
elements $S_{\alpha\beta}$ we use the relation \cite{Ab64}
\begin{eqnarray}
\label{Ecc}
 {\rm {\bf E}}(-\ve,-\xi)=-ie^{-\pi\ve}{\rm
{\bf E}}(-\ve,\xi)+it(\ve)^{-1}{\rm {\bf E}}^{\ast}(-\ve,\xi),
\end{eqnarray}
which enables us to write $\chi_{2}$ in terms of $\chi_{1}$ and
$\chi^{\ast}_{1}$. A second expression connecting $\chi_{1}$ with
$\chi_{2}$ is provided by Eq. (\ref{property}). From these relations, and
the fact that $\chi_{1}$ and $\chi^{\ast}_{1}$ are linearly independent,
we obtain
\begin{eqnarray}
 S_{11}(\ve_{n})&=&S_{22}(\ve_{n})=-r(\ve_{n}),\nonumber\\
S_{12}(\ve_{n})&=&S_{21}(\ve_{n})=-it(\ve_{n}).   \label{sm1d}
\end{eqnarray}

(ii) In order to deduce the normalization constant $C$ in (\ref{fullss})
we need to obtain the flux transmitted at $x=+\infty$ due to the
scattering state $\chi_{1n}(E|\xi ,s)$ coming from $x=-\infty$. Utilizing
the {\bf $\langle$bra$|$ket$\rangle$} notations, such that $\langle
xy|1En\rangle \equiv \chi_{1n}(E|x,y )$, we write the $x-$component of the
flux at the PC cross-section at $x_{0}$
\begin{eqnarray}
&&J_{x}(x_{0})=\Re\Bigl (\int\int dx\; dy\langle 1En|xy \rangle
\delta(x-x_{0})v_{x}\langle xy|1En \rangle\Bigr )\nonumber\\ &=&\Re\Bigl
(\langle 1En| \delta(x-x_{0})v_{x}|1En\rangle\Bigr )\nonumber\\
&=&C^2t^{2}(\ve_{n})\Re \Bigl (\int d\xi \int\ ds \Phi_{n}(s){\rm{\bf
E}}^{\ast}(-\ve_{n},\xi)\nonumber\\
&&\times[\delta(x-x_{0})v_{x}]\Phi_{n}(s){\rm{\bf E}}(-\ve_{n},\xi)\Bigr
).\label{Jx0}
\end{eqnarray}
This calculation requires the velocity operator, $v_{x}=p_{x}/m+(e{\rm
B}/2mc)y$ in the $\{\xi, s\}$ representation. Following again Ref.
\onlinecite{F87}, we have
\begin{eqnarray}
\left[\begin{array}{c}p_{x}\\y\end{array}\right]=
\left[\begin{array}{cc}\displaystyle\frac{1}{\ell}e^{\theta_{1}}\cos\phi&
\displaystyle\frac{1}{\sqrt{2}\ell}e^{-\theta_{2}}\sin\phi\\ -2\ell
e^{\theta_{1}}\sin\phi &\sqrt{2}\ell e^{-\theta_{2}}\cos\phi
\end{array}\right]
\left[\begin{array}{c}-i\partial/\partial\xi\\-i
\partial/\partial s
\end{array}\right],
\end{eqnarray}
so that $v_{x}$ is a linear combination of $\partial/\partial\xi$ and
$\partial/\partial s$. Similarly, [see Eq. (\ref{aij})], the operator $x$
in $\delta(x-x_{0})$ is a linear combination of $\xi$ and $s$. However, at
$x_{0}\rightarrow +\infty$, the term linear in $s$ can be neglected. As a
result, the terms involving $\partial /\partial s$ in Eq. (\ref{Jx0})
disappear, leading to
\begin{eqnarray}
&&J_{x}(x_{0})=C^{2}t^{2}(\varepsilon_{n})2\delta_{\rm B}a\int d\xi
\delta(a\xi -x_{0})\nonumber\\ &&\times \Re\left[{\rm{\bf
E}}^{\ast}(-\ve_{n},\xi)\left(-i\frac{\partial}{\partial\xi}\right){\rm{\bf
E}}(-\ve_{n},\xi)\right],
\end{eqnarray}
in which $a$, Eq. (\ref{p}), is the coefficient relating $x$ to
$\xi $ in (\ref{aij}). Using  the asymptotic expressions
\cite{Ab64} of ${\rm {\bf E}}(a,x)$ at $x\rightarrow +\infty$,
enables one to perform the $\xi$-integration, and with the
requirement $J_{x}(x_{0})=|S_{12}|^2$, we find
\begin{eqnarray}
C=\Bigl (\frac{1}{2\delta_{\rm B}}\Bigr )^{1/2}.
\end{eqnarray}

(iii) The matrix elements (\ref{matrixele}) can be written in the
$\{\xi,s\}$ representation
\begin{eqnarray}
M^{q}_{\beta m,\alpha n}(E)&=& \int_{-\infty}^{+\infty}dsd\xi
\chi^{\ast}_{\beta m} (E|\xi,s)\chi_{\alpha n} (E|\xi,s)\nonumber\\
&&\times\exp[iq(a\xi+bs)].
\end{eqnarray}
Here $b$ is the coefficient relating $x$ to $s$ in (\ref{aij}),
\begin{eqnarray}
b=\Bigl (\frac{1}{m\Delta_{\rm B}}\frac{\Delta^{2}_{\rm
B}-\Delta^{2}}{\Delta^{2}_{\rm B}+\delta^{2}_{\rm B}}\Bigr )^{1/2}.
\end{eqnarray}
It is zero at {\rm B}=0, increases with {\rm B}, being of the order of
the magnetic length $\ell_{\rm B}$ when $\omega_{\rm B}\simeq \Delta$, and
saturates at $\ell_{\rm B}$ when $\omega_{\rm B}\gg \Delta$. Since
$s\simeq 1$ and the SAW wavelength $2\pi/q\gg \ell_{\rm B}$, the second
term in the exponent can be neglected. Then the $s$-integration becomes
trivial, and the resulting expression is
\begin{eqnarray}
\label{MtoMm}
 M^{q}_{\beta m,\alpha n}(E)=\delta_{m,n}M_{\beta\alpha}(\ve_{n},p)
\end{eqnarray}
with
\begin{eqnarray}
\label{Mm} && M_{\beta\alpha}(\ve,p)=\frac{t^{2}(\ve)}{2\delta_{\rm B}}
\int_{-\infty}^{+\infty}d\xi \exp(ip\xi)\nonumber\\ &&\times{\rm {\bf
E}}^{\ast}(-\ve,\xi_{\beta}){\rm {\bf E}} (-\ve,\xi_{\alpha}),
  \ \ \alpha ,\beta =1,2,
\end{eqnarray}
where we have used Eq. (\ref{weber}) and the notation
$\xi_{1,2}\equiv\pm\xi$.

Inserting the results (\ref{1dme}) and (\ref{MtoMm}) into the expression
for the pumped current induced by the SAW, Eq. (\ref{JSAW}), we find
\begin{eqnarray}
&&J^{dc}_{\beta}= e\omega_{0}|{\cal A}|^{2} \sum_{\alpha n}\sum_{\beta
'\beta ''}\Bigl (S_{\beta\beta '}(\varepsilon_{n}) S^{\ast}_{\beta\beta
''}(\varepsilon_{n})\nonumber\\ &&M_{\beta '\alpha} (\varepsilon_{n},p)
M^{\ast}_{\beta ''\alpha}(\varepsilon_{n},p)-(p\rightarrow -p )\Bigr
).\label{J2}
\end{eqnarray}
We parametrize the matrix elements $M_{\alpha \beta}$ using the
representation \cite{Ab64}
\begin{eqnarray}
{\rm {\bf E}}(-\ve,\xi )&=&k^{-1/2}{\rm W}(-\ve,\xi )+ik^{1/2}{\rm
W}(-\ve,-\xi ), \nonumber\\ k^{\pm 1}&=&\sqrt{1+e^{-2\pi\ve}}\mp
e^{-\pi\ve},
\end{eqnarray}
where ${\rm W}$ are the real Weber functions. We then find
\begin{eqnarray}
M_{\stackrel{11}{22}}&=&\frac{t(\varepsilon )}{2\delta_{\rm B}}\Bigl (
K(\varepsilon ,p)\mp ir(\varepsilon )G(\varepsilon ,p)\Bigr ),\nonumber\\
M_{\stackrel{12}{21}}&=& \frac{t(\varepsilon )}{2\delta_{\rm B}}\Bigl
(2H(\varepsilon ,p) \pm t(\varepsilon )G(\varepsilon ,p)\Bigr ),
\label{M11}
\end{eqnarray}
where
\begin{eqnarray}
&&K(\ve,p)=\int_{-\infty}^{+\infty} d\xi \left ({\rm W}^{2}(-\ve,\xi
)+{\rm W}^{2}(-\ve,-\xi )\right) \cos p\xi\nonumber\\
&&G(\ve,p)=\int_{-\infty}^{+\infty} d\xi  \Bigl ({\rm W}^{2}(-\ve,-\xi
)-{\rm W}^{2}(-\ve, \xi )\Bigr )\sin p\xi,\nonumber\\
&&H(\ve,p)=\int_{-\infty}^{+\infty} d\xi  {\rm W}(-\ve,\xi ){\rm
W}(-\ve,-\xi )\cos p\xi .  \label{KGH}
\end{eqnarray}
Inserting Eqs. (\ref{M11}) and (\ref{sm1d}) into the current equation,
(\ref{J2}), we obtain that $J_{1}^{dc}=-J_{2}^{dc}\equiv J$, where $J$ is
given by Eq. (\ref{Jfinal}) above, with
\begin{eqnarray}
F(\varepsilon ,p)=t^{3}(\varepsilon )G(\varepsilon ,p)H(\varepsilon ,p).
\label{Ffinal}
\end{eqnarray}
The integrals $G$ and $H$ where calculated in \cite{LEW00}.

\acknowledgements This work was supported by the Alexander von Humboldt
Foundation (YL), the Israel Science Foundation, the German-Israeli
Foundation, and the Deutsche Forschungsgemeinschaft (PW).

\end{multicols}

\end{document}